\documentclass[namedreferences]{kluwer}
\usepackage{graphicx}

\begin{document}
\begin{article}

\begin{opening}
\title{A Method For Eclipsing Component Identification In Large Photometric Datasets}

\author{J. \surname{Devor}\email{jdevor@cfa.harvard.edu}}
\author{D. \surname{Charbonneau}}
\institute{Harvard-Smithsonian Center for Astrophysics, 60 Garden
St., Cambridge, MA 02138, USA}

\begin{abstract}
We describe an automated method for assigning the most likely
physical parameters to the components of an eclipsing binary (EB),
using only its photometric light curve and combined color. In
traditional methods (e.g. WD and EBOP) one attempts to optimize a
multi-parameter model over many iterations, so as to minimize the
chi-squared value. We suggest an alternative method, where one
selects pairs of coeval stars from a set of theoretical stellar
models, and compares their simulated light curves and combined
colors with the observations. This approach greatly reduces the EB
parameter-space over which one needs to search, and allows one to
determine the components' masses, radii and absolute magnitudes,
without spectroscopic data. We have implemented this method in an
automated program using published theoretical isochrones and
limb-darkening coefficients. Since it is easy to automate, this
method lends itself to systematic analyses of datasets consisting
of photometric time series of large numbers of stars, such as
those produced by OGLE, MACHO, TrES, HAT, and many others surveys.
\end{abstract}
\end{opening}

\section{Introduction}

Eclipsing double-lined spectroscopic binaries provide the only
method by which both the masses and radii of stars can be
estimated without having to resolve spatially the binary or rely
on astrophysical assumptions. Despite the large variety of models
and parameter-fitting implementations (e.g. WD and EBOP), their
underlying methodology is essentially the same. Photometric data
provides the light curve of the EB, and spectroscopic data provide
the radial velocities of its components. The depth and shape of
the light curve eclipses constrain the components' brightness and
fractional radii, while the radial velocity sets the length scale
of the system. In order to characterize fully the components of
the binary, one needs to combine all of this information.
Unfortunately, only a small fraction of all binaries eclipse, and
spectroscopy with sufficient resolution can be performed only for
bright stars. The intersection of these two groups leaves a
pitifully small number of stars.

In the past decade, there has been a dramatic growth in the number
of stars with high-quality, multi-epoch, photometric data. This
has been due to major advances in both CCD detectors and the
implementation of image-difference analysis techniques
\cite{Crotts92,Alard98, Alard00}, which enables simultaneous
photometric measurements of tens of thousands of stars in a single
exposure. Today, there are many millions of light curves available
from a variety of surveys, such as OGLE, \inlinecite{Udalski94};
MACHO, \inlinecite{Alcock98}; TrES, \inlinecite{Alonso04}; and
HAT, \inlinecite{Bakos04}. But there has not been a corresponding
growth in the quantity of spectroscopic data, nor is this likely
to occur in the near future. Thus, the number of
fully-characterized EBs has not grown significantly. In recent
years there has been a growing effort to mine the wealth of
available photometric data, by employing simplified EB models in
the absence of spectroscopic observations (\opencite{Wyithe01},
\citeyear{Wyithe02}; \opencite{Devor04}, \citeyear{Devor05}).

In this paper we present a novel approach, which utilizes
theoretical models of stellar properties to estimate the orbital
parameters as well as the masses, radii, and absolute magnitudes
of the stars, while requiring \emph{only} a photometric light
curve and an estimate of the binary's combined color. This
approach can be used to characterize quickly large numbers of
eclipsing binaries, however it is not sufficient to improve
stellar models since underlying isochrones must be assumed. We
have created two implementations of this idea. The first, which we
have named MECI-express, and is described in section
\ref{secMECIexp}, is a ``quick and dirty'' program that is
designed as a simple extension to the Detached Eclipsing Binary
Light curve (DEBiL) fitter (\opencite{Devor04},
\citeyear{Devor05}). The second, which we have named MECI, and is
described in section \ref{secMECI}, is considerably more accurate,
but also more computationally demanding. The source code for both
MECI-express and MECI will be provided upon request.

\section{Express Method for Eclipsing Component Identification (MECI-express)}
\label{secMECIexp}

The primary application of MECI-express is to identify the stellar
components of a given EB. It operates after a conventional EB
model-fitting program has already analyzed the given EB's light
curve. In our implementation, we chose to employ DEBiL
(\opencite{Devor04}, \citeyear{Devor05}) since it is simple, fast,
and fully automated. The fitted parameters are the orbital period
(P), the apparent magnitudes ($mag_{1,2}$), and the fractional
radii ($r_{1,2}$) of the binary components. A fractional radius is
defined as the radius ($R_{1,2}$) divided by the sum of the
components' semimajor axes (a). In MECI-express we iterate through
a large group of MK spectral type pairings, to each of which we
associate typical stellar parameters \cite{Cox00}. These stellar
parameters are the masses ($M_{1,2}$), the radii ($R_{1,2}$), and
the absolute magnitudes ($Mag_{1,2}$) of the binary components. If
the assumed values of the stellar parameters match the true
values, then the stellar and fitted parameters should obey to the
following equations:

\begin{eqnarray}
\frac{4\pi^2R_1^3}{G(M_1 + M_2)} &=& P^2r_1^3 \label{eq_r1}\\
\frac{4\pi^2R_2^3}{G(M_1 + M_2)} &=& P^2r_2^3 \label{eq_r2}\\
Mag_1 - Mag_2 &=& mag_1 - mag_2 \label{eqMag}
\end{eqnarray}

We also may have additional constraints from the observed
out-of-eclipse combined colors of the system. For example, in the
case of OGLE~II targets, we have the estimated V-I color:

\begin{equation}
Mag_V - Mag_I = mag_V - mag_I \label{eqColor}
\end{equation}

\begin{figure}[H]
\tabcapfont
\centerline{%
\begin{tabular}{ccc}
\includegraphics[width=1.75in]{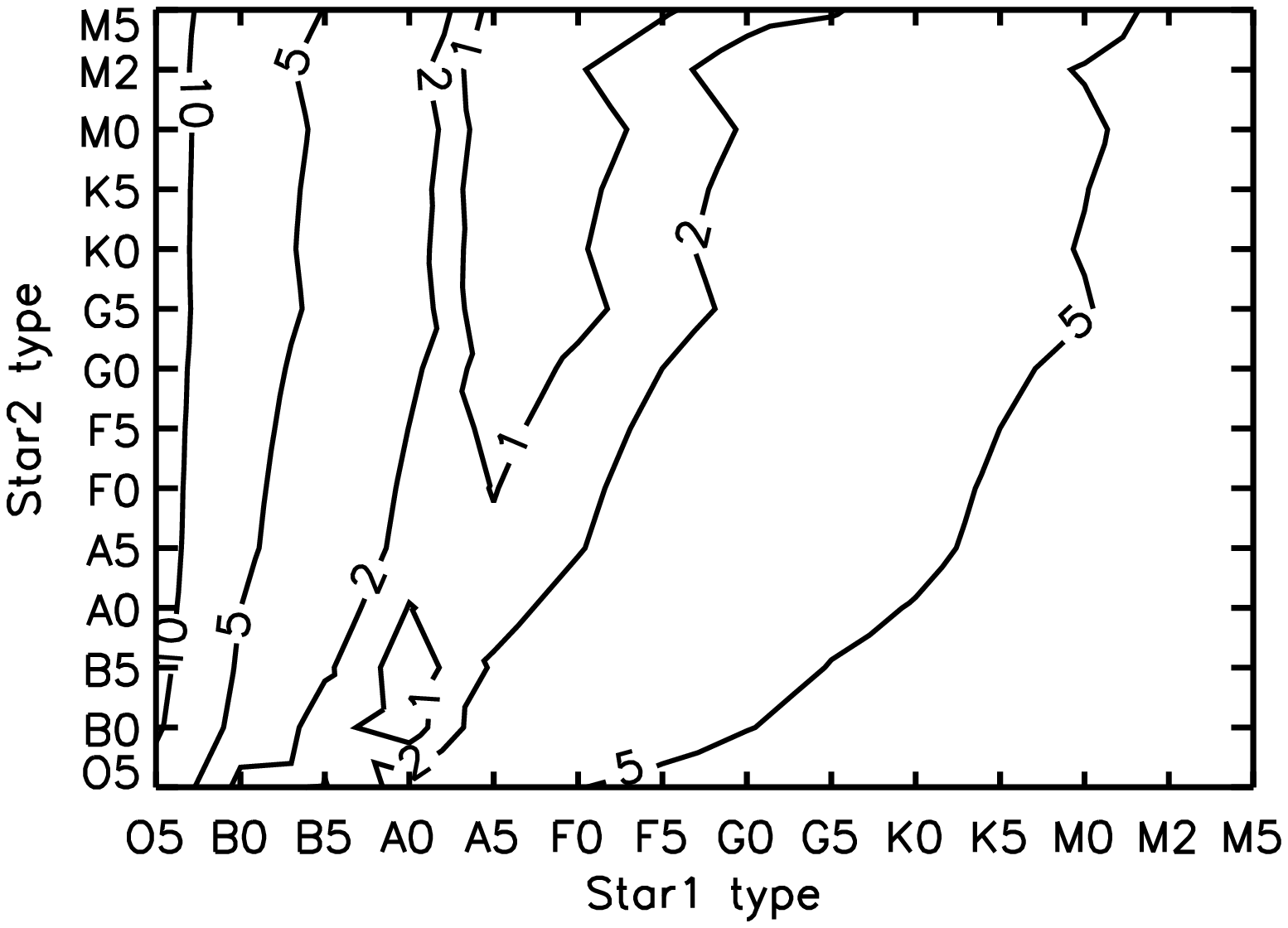} &
\includegraphics[width=1.75in]{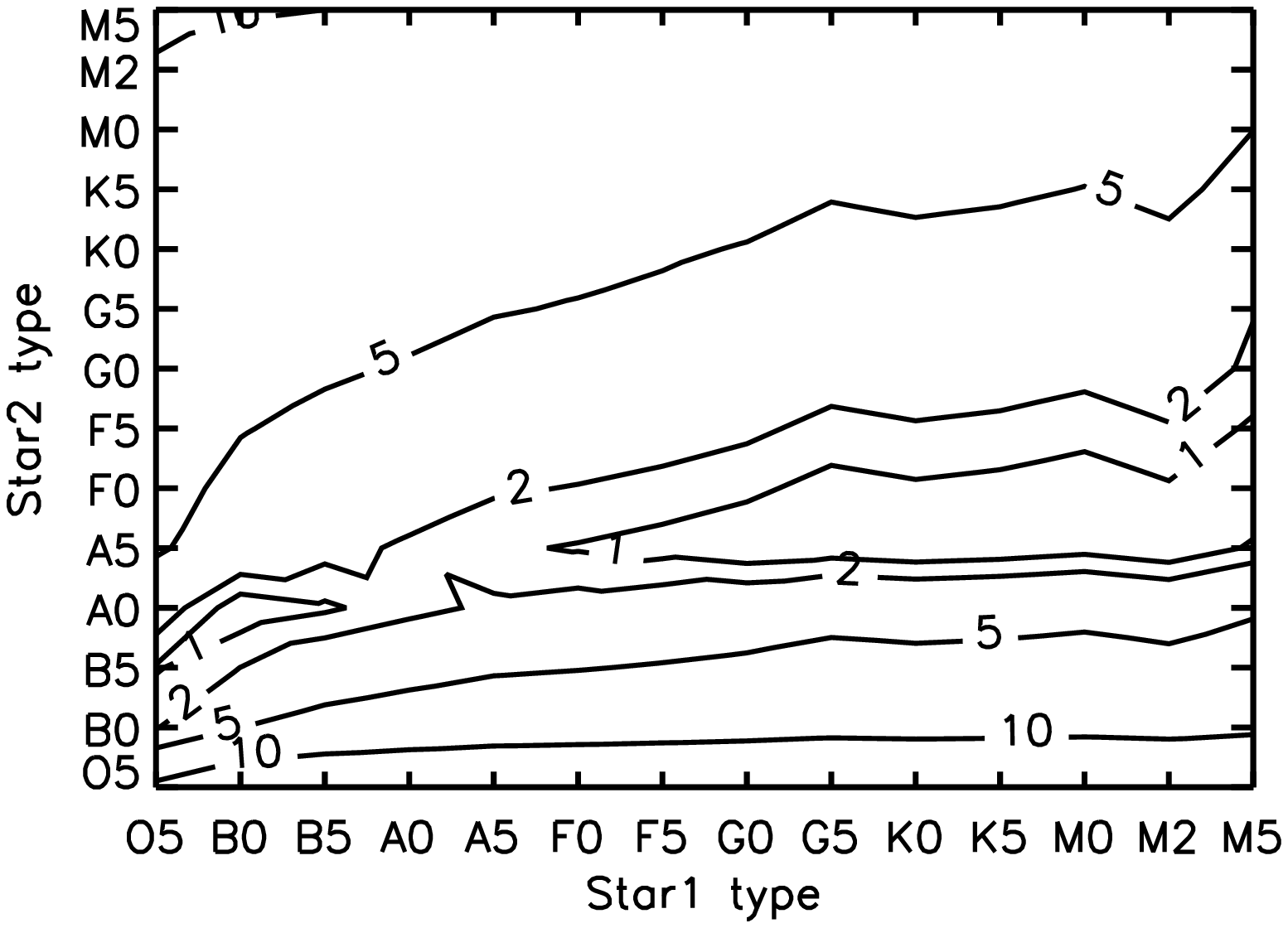} &
\includegraphics[width=1.75in]{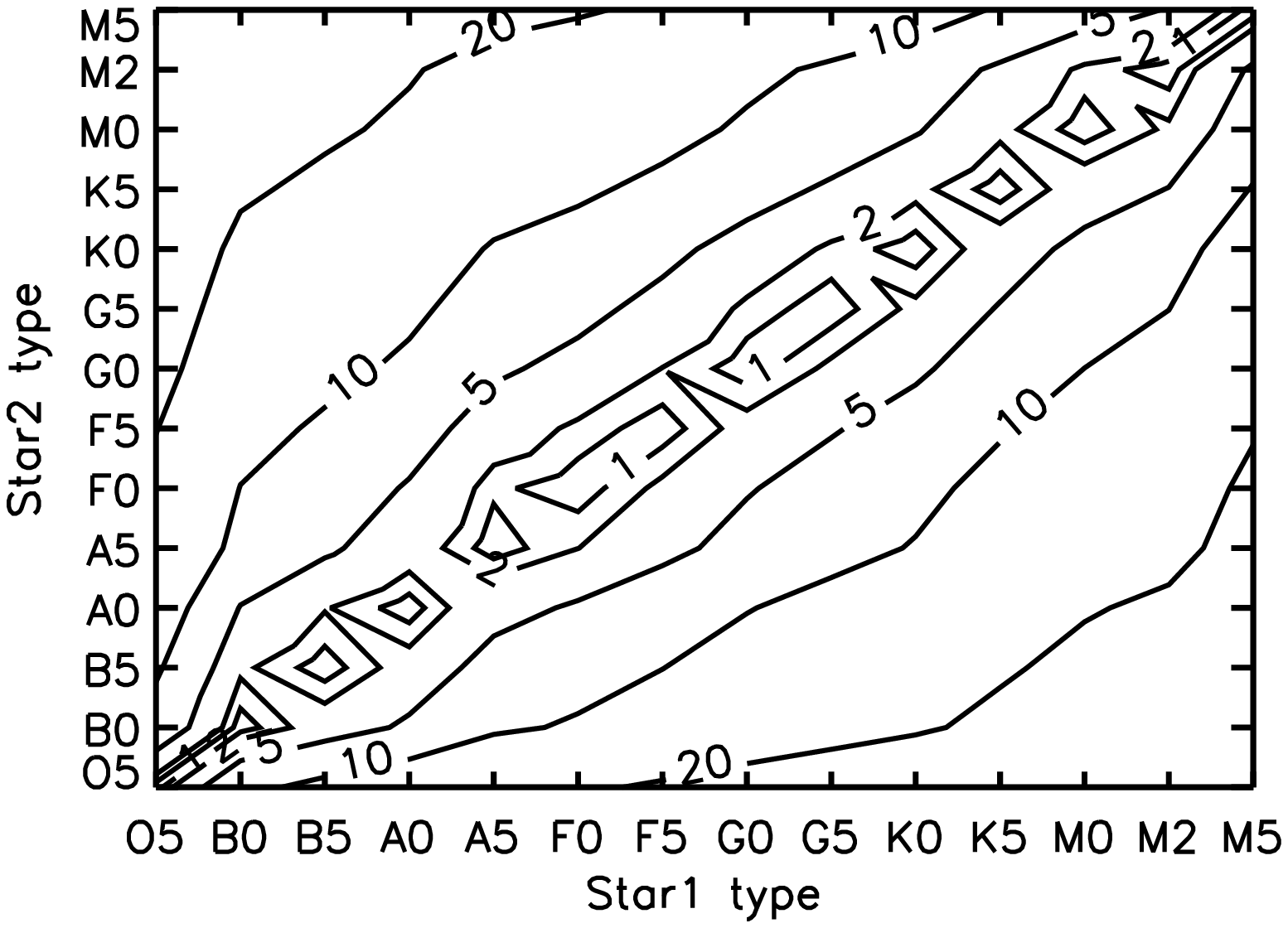}\\
a. Constraints from (eq. \ref{eq_r1}) & b. Constraints from (eq.
\ref{eq_r2}) & c. Constraints from (eq. \ref{eqMag})
\end{tabular}}
\caption{Contour plots of the absolute difference between the
left-hand-side and the right-hand-side of each equation, divided
by its uncertainty, as applied to the WW Camelopardalis light
curve \cite{Lacy02}. Adding these results in quadrature, produces
the likelihood plot shown in Figure~\ref{fig2}.a.}
\label{fig1}
\end{figure}

\begin{figure}[H]
\tabcapfont
\centerline{%
\begin{tabular}{c@{\hspace{2pc}}c}
\includegraphics[width=2.4in]{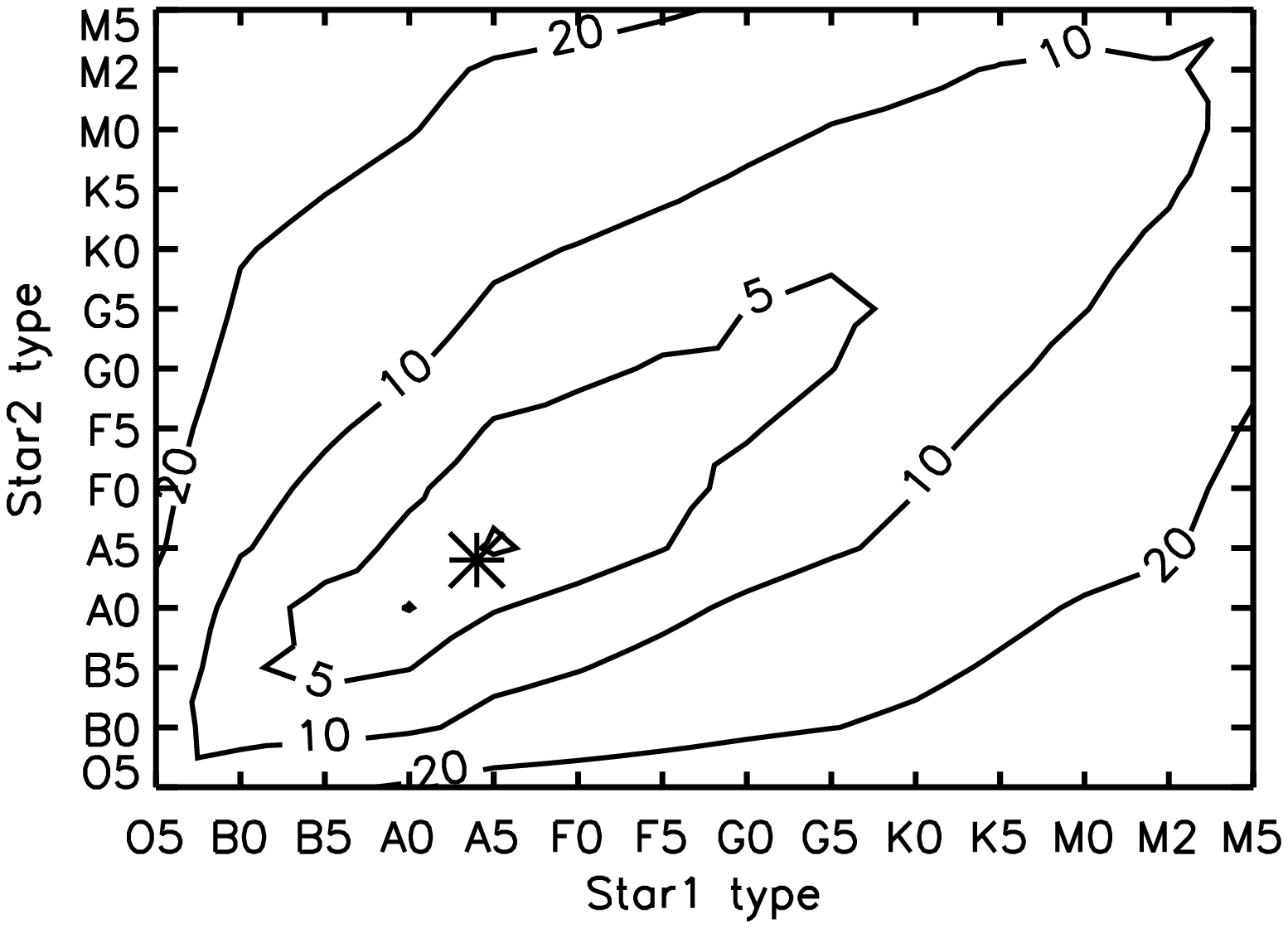} &
\includegraphics[width=2.4in]{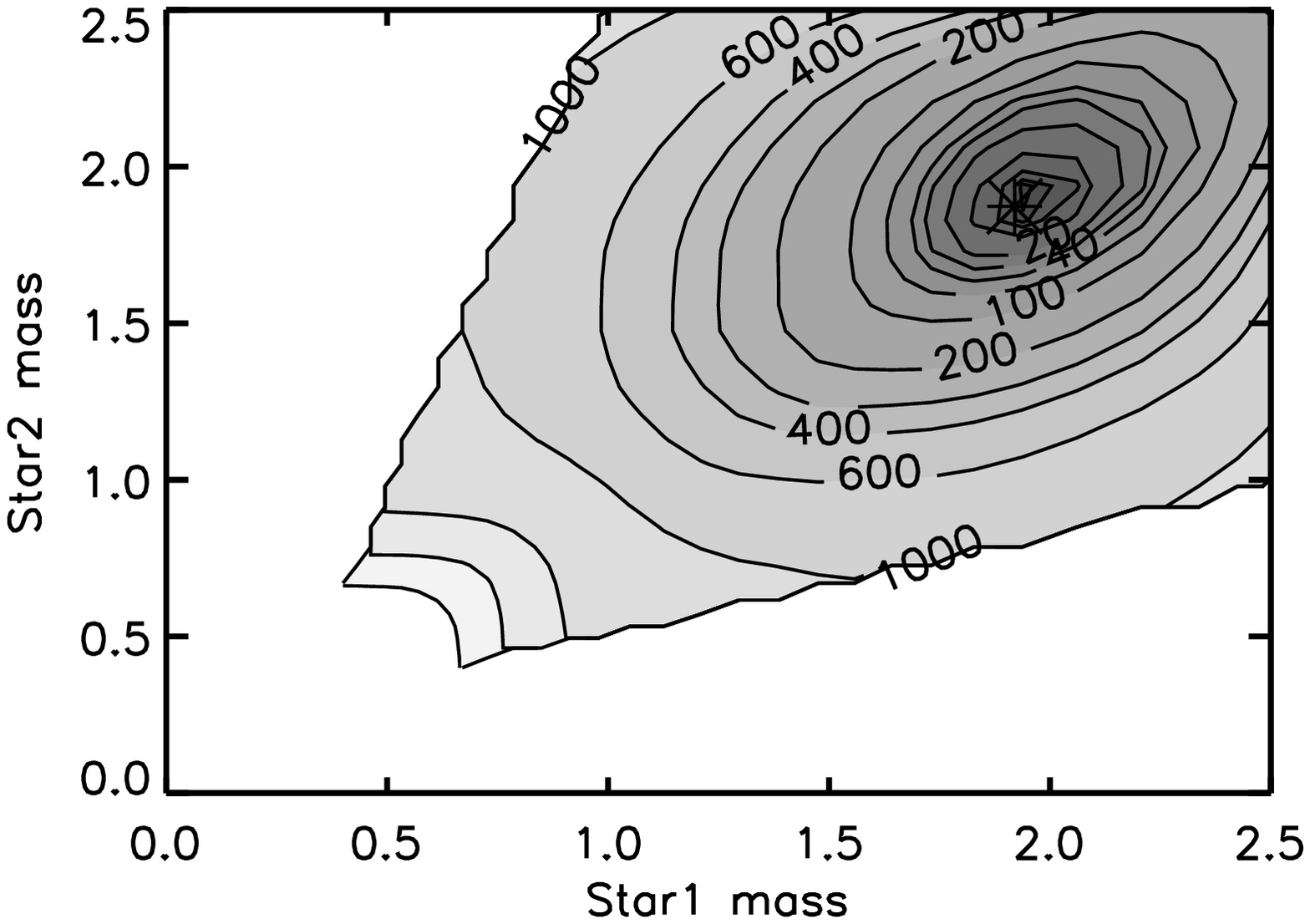} \\
a. MECI-express likelihood plot & b. MECI likelihood plot (age =
0.6 Gyr)
\end{tabular}}
\caption{A comparison of the MECI-express (left) and MECI (right)
likelihood contour plots for WW Camelopardalis. The value of the
contours are described in the body of the text. The asterisk marks
the solution of \inlinecite{Lacy02}.}
\label{fig2}
\end{figure}

We assume that the color has been corrected for reddening and that
no systematic errors are present, so any inequalities would be due
to an incorrect choice for the component pairing. The likelihood
of each pairing is assessed by calculating the difference between
the left-hand-side (stellar parameters) and right-hand-side
(fitted parameters) of each equation. These differences are
divided by their uncertainties, and added in quadrature. The
pairing with the smallest sum is deemed the most likely pairing.
For each given EB light curve, MECI-express returns the list of
the top ranked (most likely) binary pairings, with their
corresponding sums. MECI-express can also be used to create a
contour plot of the probability distribution for all pairings. We
illustrate an example of individual MECI-express components in
Figures~\ref{fig1}.a-c, which are then combined to create the
result shown in Figure~\ref{fig2}.a.

\section{Method for Eclipsing Component Identification (MECI)}
\label{secMECI}

MECI was developed to improve significantly upon the accuracy of
MECI-express (see Table 1). This was done as follows: We replaced
the use of spectral types with the more fundamental (and
continuous) quantities of mass and age. Furthermore, in MECI we
assume that the two binary components are coeval, thus replacing
the 2-dimensional spectral type - spectral type grid, with a
3-dimensional mass-mass-age grid. Finally, we no longer rely on
parameter fits of the components' apparent magnitudes and
fractional radii directly from the light curve, which are often
very uncertain, nor do we assume constant limb-darkening
coefficients. Instead, we interpolate these values for the given
mass-mass-age pairing, from precalculated tables [Yonsei-Yale
isochrones \cite{Kim02} ; ATLAS limb-darkening coefficients
\cite{Kurucz92}, used when $T_{eff} \geq 10000K$ or $\log g \leq
3.5$  ; PHOENIX limb-darkening coefficients (\opencite{Claret98},
\citeyear{Claret00}), used when $T_{eff} < 10000K$ and $\log g >
3.5$]. Thus, by assuming the masses ($M_{1,2}$) of the EB
components and the system's age, we can look-up the radii
($R_{1,2}$), the absolute magnitudes ($Mag_{1,2}$), and the
limb-darkening coefficients for the binary components. We then use
these values, as well as the observationally-determined period (P)
and combined magnitude out of eclipse ($mag_{comb}$), to calculate
the apparent magnitudes ($mag_{1,2}$) and factional radii
($r_{1,2}$) of the EB components, as follows:

\begin{eqnarray}
mag_1 &=& mag_{comb} + 2.5\log \left[1 + 10^{-0.4(Mag_2 - Mag_1)}\right]\\
mag_2 &=& mag_1 + (Mag_2 - Mag_1)\\
a &=& [G(M_1 + M_2)(P/2\pi)^2]^{1/3} \simeq \nonumber\\
  & & 4.206 R_{\odot}(M_1/M_{\odot} + M_2/M_{\odot})^{1/3} P^{2/3}_{day}\\
r_{1,2} &=& R_{1,2} / a
\end{eqnarray}

Besides the epochs of eclipses, which can be determined directly
from the EB light curve, there are only two additional parameters
required for us to simulate the light curves of the given pairing:
the orbital eccentricity (e) and inclination (i). For binaries
with short periods ($\lesssim$~2~days) and a secondary eclipse
precisely half an orbit after the primary eclipse, it is
reasonable to assume a circular orbit ($e=0$). Otherwise, one
should use the eccentricity derived by an EB model-fitting program
(we use DEBiL). Finding the inclination robustly is more
difficult. We employ a bracket search \cite{Press92}, which
returns the inclination that produces the best resulting fit.

To summarize, for every combination of component masses and system
age of an EB, we can look-up, calculate, or fit all the parameters
needed to simulate its light curve (P, limb-darkening
coefficients, mag$_{1,2}$, r$_{1,2}$, epochs of eclipses, e, i),
as well as its apparent combined color. We systematically iterate
through many such combinations. For each one we compare the
expected light curve with the observations, and calculate the
reduced chi-squared value ($\chi^2_\nu$). We also compare each
observed color ($O_c \pm \epsilon_c$) with its calculated value
($C_c$), and combine them by defining: $score\equiv(w\chi^2_\nu +
\sum_{c=1}^N [(O_c - C_c)/\epsilon_c]^2)/(w+N)$. Where $w$ is the
$\chi^2_\nu$ information weighting. We use $w=1$, and assume that
the smaller the score, the more likely it is that we have chosen
the correct binary pairing. One can visualized this result using a
series of $score(M_1, M_2)$ contour plots, each with a constant
age (e.g. Figure~\ref{fig2}.b).

\section{Conclusions}
\label{secConclusions}

We have described a novel method for identifying an EB's
components using only its photometric light curve and combined
color. By utilizing theoretical isochrones and limb-darkening
coefficients, this method greatly reduces the EB parameter-space
over which one needs to search. This approach seeks to estimate
the masses, radii and absolute magnitudes of the components,
without spectroscopic data. We described two implementations of
this method, MECI-express and MECI, which enable systematic
analyses of datasets consisting of photometric time series of
large numbers of stars, such as those produced by OGLE, MACHO,
TrES, HAT, and many others. Such techniques are expected to grow
in importance with the next generation surveys, such as Pan-STARRS
\cite{Kaiser02} and LSST \cite{Tyson02}.

\newpage

\begin{table}
\caption{A comparison of the results produced by MECI-express,
MECI, and conventional analyses with their uncertainties (Lacy et
al., 2000, 2002, 2003). The square brackets with numerical values
indicate the deviation of our results from those of the
conventional approach.}
\begin{tabular}{c|cc|ccc|ccc}
\hline & \multicolumn{2}{c|}{MECI-express} &
\multicolumn{3}{c|}{MECI} & \multicolumn{3}{c}{Lacy et al. (2000, 2002, 2003)} \\
Parameter & Mass 1       & Mass 2        & Mass 1        & Mass 2        & Age   & Mass 1        & Mass 2        & Age   \\
      &[$M_{\odot}$] & [$M_{\odot}$] & [$M_{\odot}$] & [$M_{\odot}$] & [Gyr] & [$M_{\odot}$] & [$M_{\odot}$] & [Gyr] \\
\hline
FS Mon & 2.9 (A0) & 2.0 (A5) & 1.62     & 1.52     &  1.4  & 1.632      & 1.462      & 1.6      \\
       & [77.7\%] & [36.8\%] & [0.6\%] & [4.1\%] & [0.2] & $\pm$0.012 & $\pm$0.010 & $\pm$0.3 \\
WW Cam & 2.0 (A5) & 2.0 (A5) & 1.97     & 1.89     &  0.5  & 1.920      & 1.873      & 0.5      \\
       & [4.2\%]  & [6.8\%]  & [2.8\%] & [1.0\%] & [0.0] & $\pm$0.013 & $\pm$0.018 & $\pm$0.1   \\
BP Vul & 2.0 (A5) & 1.6 (F0) & 1.77     & 1.48     &  0.7  & 1.737      & 1.408      & 1.0      \\
       & [15.1\%] & [13.6\%] & [2.1\%] & [5.4\%] & [0.3] & $\pm$0.015 & $\pm$0.009 & $\pm$0.2 \\
\hline
\end{tabular}
\end{table}

We are grateful to Guillermo Torres for many helpful
conversations.

{}

\end{article}
\end{document}